\documentclass{mn2e}
\usepackage{psfig}
\usepackage{mnras_cite}
\newlength{\tskip}
\newlength{\colwidth}

\newcommand{\beq}{\begin{equation}}
\newcommand{\eeq}{\end{equation}}
\newcommand{\beqa}{\begin{eqnarray}}
\newcommand{\eeqa}{\end{eqnarray}}

\newcommand{\rad}{r}    

\begin{document}

\title[GWB due to Cosmological Neutron Star-White Dwarf Binaries]{Gravitational Wave Background of Neutron Star-White Dwarf Binaries}
\author[Cooray]{Asantha Cooray\\
Theoretical Astrophysics, California Institute of Technology, 
MS 130-33, Pasadena, CA 91125, USA}

\maketitle

\begin{abstract}
We discuss the stochastic background of gravitational waves from ultra compact neutron star-white dwarf (NS-WD) binaries
at cosmological distances. Under the assumption that accreting neutron stars and donor white dwarf stars  form
most of the low mass X-ray binaries (LMXBs),  our calculation makes use of recent results related to 
the luminosity function determined from X-ray observations. 
Even after accounting for detached NS-WD binaries not captured in X-ray data,
the NS-WD background is at least an order of magnitude below that due to extragalactic white dwarf-white dwarf binaries and  
below the detectability level of the Laser Interferometer Space Antenna (LISA)
at frequencies between 10$^{-5}$ Hz and 10$^{-1}$ Hz. While the extragalactic background is unlikely to be detected, we suggest 
that around one to ten
galactic NS-WD binaries may be resolved with LISA such that their positions are determined to an accuracy of
several degrees on the sky. 
\end{abstract}


\section{Introduction}
The planned Laser Interferometer Space Antenna (LISA)\footnote{http://lisa.jpl.nasa.gov and http://sci.eas.int/home/lisa/} 
will open up the frequency band between 
 10$^{-5}$ Hz and 10$^{-1}$ Hz to the gravitational wave radiation. 
At frequencies below 10$^{-5}$ Hz, merging super massive black holes at cosmological distances contribute 
(Jaffe \& Backer 2003; Wyithe \& Loeb 2003), 
while at frequencies above 10$^{-1}$ Hz, neutron star-neutron star binaries and stellar mass black holes are
expected to be the dominant sources of gravitational waves (Belczynski et al. 2002).
The range between  10$^{-5}$ Hz and 10$^{-1}$ Hz is ideal for the detection of gravitational waves from
compact binaries associated with close white dwarf binaries (CWDBs) involving another white dwarf, a main sequence star or
a neutron star (Hils et al. 1990; Farmer \& Phinney 2003; Nelemans et al. 2001; Cooray et al. 2004).

In addition to observing the coalescence of massive black hole binaries (Haehnelt 1994), 
at frequencies above 10$^{-3}$ Hz, LISA will spectrally resolve most galactic CWDBs (Cornish \& Larson 2003). Estimates
based on population synthesis calculations suggest the number of resolved binaries
 will be around 5000, with a substantial fraction related to WD-WD binaries
in the galaxy, including Am CVn systems (e.g., Nelemans et al. 2001).
At frequencies below 10$^{-3}$ Hz, the stochastic background formed by unresolved
galactic binaries is expected to be significant and be larger in amplitude when compared to both the
LISA detector noise and the extragalactic background formed by merging
white dwarfs (Hiscock et al. 2000; Hils, Bender \& Webbink 1990; Kosenko \& Postnov 1998; Nelemans et al. 2001). 
As discussed in Seto \& Cooray (2004; also, Ungarelli \& Vecchio 2001), 
this galactic background  of gravitational waves is highly anisotropic as
it is mostly concentrated towards the galactic plane and the bulge. With integration times over a year, LISA is expected to measure
low order multipoles of this galactic binary distribution. This allows a mechanism to quantify the galactic background independent of
any other isotropic background component at the same gravitational wave frequencies. 
Once the galactic background is modeled and removed, under the assumption that the spatial anisotropy of any extragalactic
background related to compact binaries is negligible, one can search for a component 
from cosmological distances. While one expects an isotropic background from inflation (see, Maggiore 2000), the extragalactic background
related to compact binaries is expected to dominate. Any measurement of the  gravitational wave background 
amplitude related to compact binaries would be helpful in understanding their cosmological evolution.

Here, we consider the extragalactic compact binary background and its detectability with LISA
and primarily focus on the background related to neutron star-white dwarf binaries;
Previous studies have already estimated the extragalactic background related to white dwarfs (orbiting other
white dwarfs and main sequence stars). These estimates have mostly relied on numerical population synthesis models based on 
codes such as SEBA (Portegies Zwart \& Yungelson 1998 and used by Schneider et al. 2001) and
BSE (Hurley et al. 2002 used by Farmer \& Phinney 2003).  While these estimates are more
likely to capture the WD-WD and WD-main sequence background more accurately, we avoid the approach related to binary synthesis
models as the physics related to the formation of neutron stars is expected to be
more uncertain than those involved with the formation of white dwarfs.

In order to calculate the background, we make use of recent results in the literature related to
the presence of low mass X-ray binaries (LMXBs) in nearby galaxies. Following Bildsten \& Deloye (2004) and
various other suggestions in the literature (e.g., Belczynski \& Taam 2004; Bildsten 2002; Podsiadlowski et al. 2002; see reviews
by Verbunt 1993; Bhattacharya \& van den Heuvel 1991), we make the 
identification that an individual LMXB is an accreting neutron star from a donor white dwarf star with mass at or below 0.1 m$_{\sun}$. 
This identification allows us to present a calculation  of the gravitational wave background based solely on observational
data, mainly using the distribution of LMXBs in each galaxy (e.g., Gilfanov 2004; Kim \& Fabbiano 2004), and the distribution of
galaxies out to $z \sim 1$ or more based on the K-band luminosity function (Drory et al. 2003);
We make use of the observationally derived result that the LMXB distribution in each galaxy can be described by a
universal luminosity function with the normalization given by the total stellar mass content.

The LMXBs, however, only capture the fraction of NS-WD binaries
that are interacting; Thus, our  approach based only on X-ray data allows us to derive a lower limit to the NS-WD background. 
We partly correct for the fraction of detached binaries based on observed statistics of binaries both in galactic globular 
clusters and the field and using certain scaling arguments from the literature related to the evolution of these binaries
including the end point of mass transfer where neutron stars are spun up to millisecond pulsars (see, Lorimer 2001 for a review).  

Though the estimate related to the total NS-WD background is rather uncertain, 
in general, our calculations suggest that the NS-WD gravitational wave background is roughly an order of magnitude smaller than those
due to WD-WD binaries. In comparison, we find that the background is unlikely to be a significant source of signal 
for LISA as it is roughly two orders of magnitude below the LISA instrumental noise at frequencies near and above a mHz. 
Our conclusions are consistent with a recent order of magnitude estimate on the NS-WD gravitational wave background at
frequencies above a mHz where the signal was estimated to be at a level an order of magnitude below the
LISA noise (Kim et al. 2004). While the analysis by Kim et al. (2004) involved statistics of three known NS-WD binaries
that will merge within a Hubble time, we present an independent estimate that relies on the observed statistics of X-ray binaries.

The discussion is organized as follows: in the next section, we discuss 
the contribution to the gravitational wave background from NS-WD binaries
under the assumption that LMXBs are binaries whose X-ray luminosity is powered
by accretion. Making use of calculations by Bildsten \& Deloye (2004) to relate the accretion luminosity
to secondary WD mass and the orbital period of the binary, we use observationally derived results on the distribution of
LMXBs to calculate the stochastic background of gravitational waves. Since not all NS-WD binaries are not included in the calculation,
we make a rather uncertain order of magnitude correction to estimate the total background related to both interacting and
detached NS-WD binaries. We find a low background signal, when compared to LISA noise.
We also consider the detectability of galactic NS-WD binaries and
conclude with a discussion and a summary in Section~3. We assume a flat cosmological model with a cosmological constant and 
consistent with recent cosmological data including WMAP.

\section{Gravitational Waves from NS-WD Binaries}

The gravitational wave luminosity for a binary in a circular orbit is given by the Peters \& Mathews (1963)
formula:
\begin{equation}
L^{\rm circ}_{\rm GW} = 2.16 \times 10^{45} {\rm ergs\; s^{-1}} \left(\frac{m_1}{m_{\sun}}\right)^{10/3} \frac{q^2}{(1+q)^{2/3}} P_{\rm orb}^{-10/3} \, ,
\label{eqn:circ}
\end{equation}
where the orbital period, $P_{\rm orb}$ is in seconds and $q\equiv m_2/m_1$ is
the ratio of masses of the binary components with individual masses $m_1$ and $m_2$. In the circular orbit case,
the gravitational wave emission is at a frequency $f_{\rm GW}=2/P_{\rm orb}$ given by
\begin{equation}
f_{\rm GW} = 3.5 \times 10^{-3} {\rm Hz} \left(\frac{m_{\rm tot}}{0.9 m_{\sun}}\right)^{1/2} \left(\frac{a}{10^5 {\rm km}}\right)^{-3/2} \, ,
\end{equation}
where $a$ is the semi-major distance of the orbit and time variation of this frequency is $\dot{f} = 96 \pi^{8/3}/5 m_c^{5/3} f^{11/3}$,
where $m_c$ is the chirp mass given by $m_c = (m_1m_2)^{3/5}(m_1+m_2)^{-1/5}$.

 In general, with an orbital eccentricity, $e$,  the gravitational wave emission
is at all integer harmonics of the orbital frequency such that $f_{\rm GW}=n/P_{\rm orb}$, where $n > 2$. In this general case,
the specific luminosity in gravitational waves can be written in terms of the circular luminosity, 
$L_{\rm GW}^f=\sum_n L^{\rm circ}g(n,e)\delta(f-n/P_{\rm orb})$, where
$g(n,e)$ accounts for the luminosity distribution as a function of the orbital eccentricity and the harmonic number 
(see, Peters \& Mathews 1963). 


Here we attempt to calculate the background independent of population synthesis models.
The basis for this calculation comes from the interpretation of LMXBs as
accreting neutron stars from donor white dwarf stars (Bildsten \& Deloye 2004). This interpretation is consistent with
a variety of observational results and the association of LMXBs with millisecond pulsars (e.g., see Lorimer 2001 and references therein).
We relate the accretion luminosity to the orbital period and donor mass.
The orbital period and donor mass distribution of NS-WD binaries, in each galaxy, then, follows from X-ray observations.
We describe the X-ray luminosity function following the description given in 
Kim \& Fabbiano (2004; also, Gilfanov 2004)
and write
\begin{eqnarray}
\frac{dN}{dL_x} = \left\{\begin{array}{ll}
A L_x^{-\alpha} & L_{\rm min} \leq L_x < L_{\rm cut}\\
A L_x^{-\beta} & L_{\rm cut} \leq L_x < L_{\rm max}\\
\end{array}\right.\,
\label{eqn:xray}
\end{eqnarray}
where $\alpha$ and $\beta$ take numerical values of  $1.8 \pm 0.2$ and $2.7 \pm 0.5$
when $L_{\rm cut} = (4.8 \pm 1.2) \times 10^{38}$ ergs s$^{-1}$. The normalization constant $A$ is
set under the observationally derived result that the total X-ray luminosity, $L_x^{\rm tot}\equiv\int L_x dN/dL_x\; dL_x$,
scales with the stellar mass as determined by the K-band luminosity, L$_K$, of the galaxy such that
when $L_{\rm min} = 10^{37}$ ergs s$^{-1}$, $L_x^{\rm tot} = (0.20 \pm 0.08) \times 10^{30}(L_K/L_{K\sun})$ ergs s$^{-1}$
where $L_{K\sun}$ corresponds to an absolute magnitude M$_{K\sun}$ of 3.33.
Note that $L_{\rm cut}$ corresponds roughly to the Eddington limit of an accreting neutron star (Bildsten \& Deloye 2004).

\begin{figure}
\centerline{\psfig{file=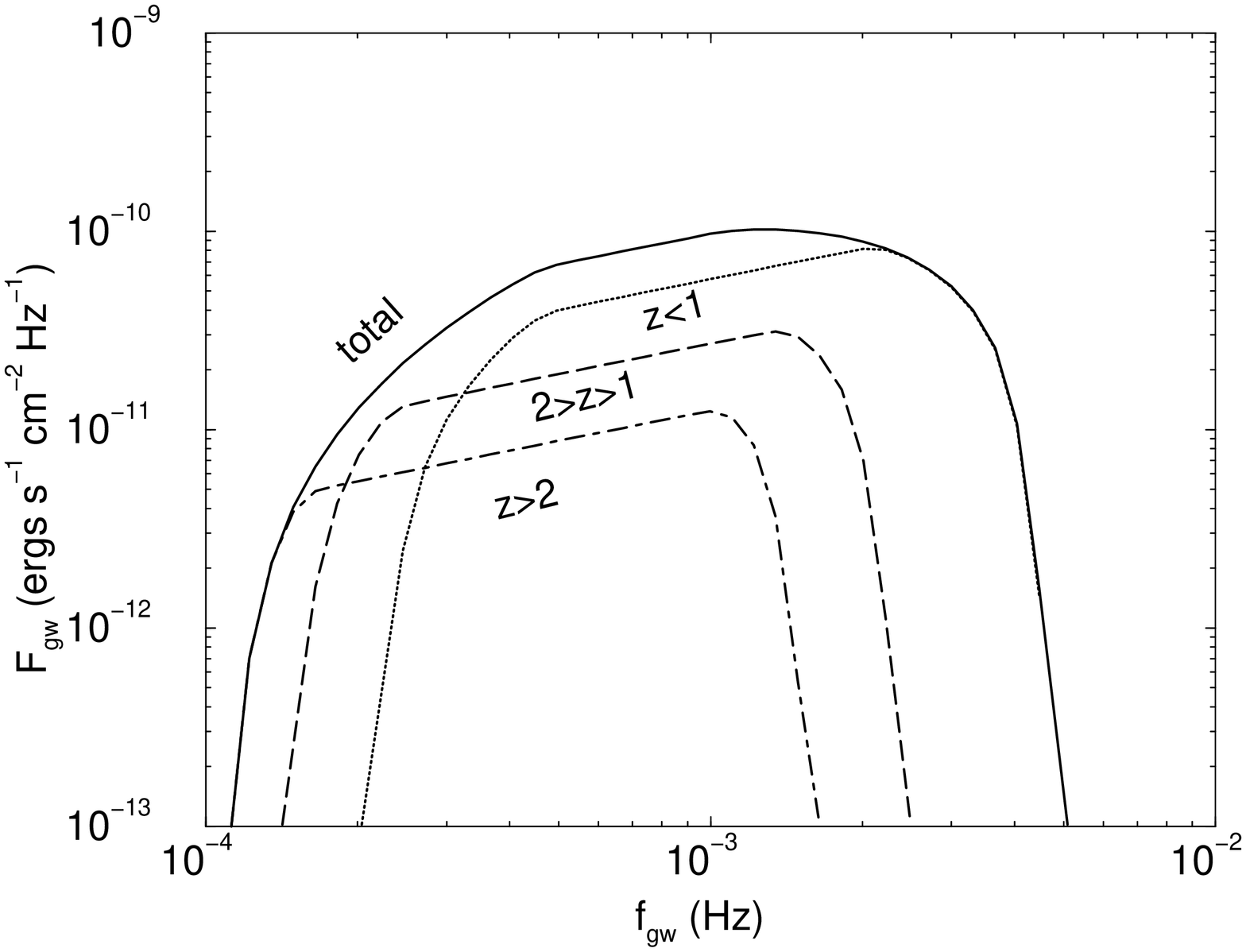,width=3.8in,angle=-90}}
\centerline{\psfig{file=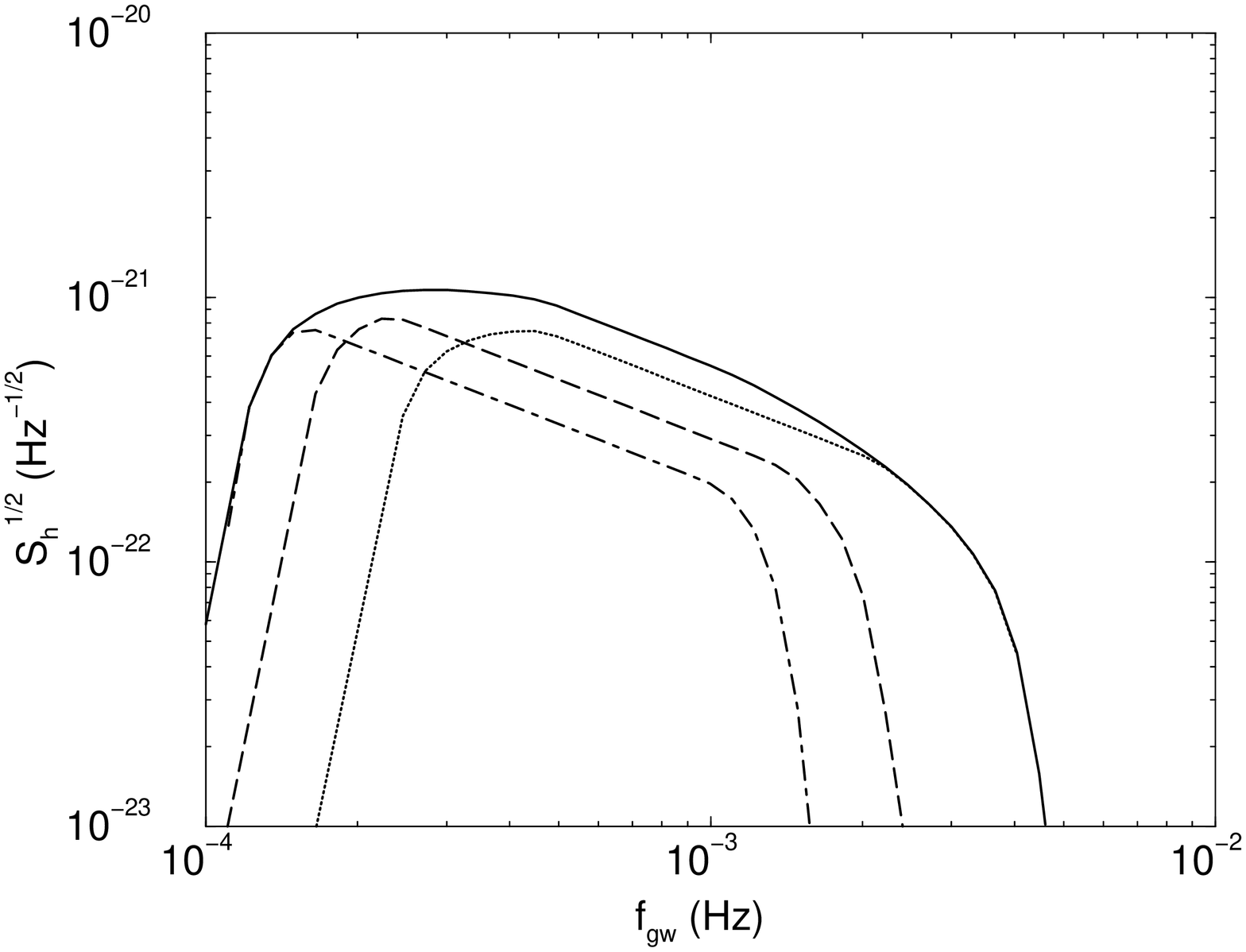,width=3.8in,angle=-90}}
\caption{The gravitational background related to NS-WD binaries based on data related to LMXBs. The background is calculated for the
fiducial parameters of the LMXB X-ray luminosity function, the K-band galaxy luminosity function, and the X-ray luminosity-secondary mass and
period relations from the literature. The background is shown in terms of the specific flux, in units of ergs s$^{-1}$ cm$^{-1}$ Hz$^{-1}$ (top panel),
and in terms of the spectral flux density, in units of Hz$^{-1/2}$ (bottom panel), as a function of the gravitational wave frequency. The total 
background signal is shown as a solid line, while, for reference, we also divide the contribution to redshift intervals of unity.}
\label{fig:gw}
\end{figure}

We extend the LMXB distribution to cosmological distances using the K-band luminosity function of field galaxies 
out to a z of 1.2 derived by Drory et al. (2003) using the Munich Near-Infrared Cluster Survey (MUNICS). Their results
are summarized in the usual Schechter (1976) form for the luminosity function such that
\begin{equation}
\Phi(L_K) dL_K = \frac{\Phi_\star}{L_K^\star}\left(\frac{L_K}{L_K^\star}\right)^\gamma \exp\left(-\frac{L_K}{L_K^\star}\right)
dL_K\, ,
\end{equation}
where $\gamma$ is the faint-end slope of the luminosity function and $\Phi_\star$ is the normalization at the characteristic
luminosity of $L_K^\star$, with the corresponding magnitude $M_K^\star$. 
The redshift evolution of the luminosity distribution is written as
\begin{eqnarray}
\Phi_\star(z) &=& \Phi_\star(0)(1+\mu z) \nonumber \\
M_K^\star(z) &=& M_K^\star(0) + \nu z \, .
\end{eqnarray}
We make use of the numerical values derived in Drory et al. (2003; see also, Kochanek et al. 2001) 
such that $\Phi_\star(z=0)=(1.16 \pm 0.10) \times 10^{-2} h^3$ Mpc$^{-3}$, $M_\star(z=0) = -23.39 + 5\log h \pm 0.05$,
and $\gamma(z) = \gamma = -1.09 \pm 0.06$. To describe evolution, we use $\mu = -0.25 \pm 0.05$ and $\nu = -0.53 \pm 0.07$,
as determined using maximum likelihood fits to the measured luminosity functions. All 
uncertainties here and in the LMXB X-ray luminosity function (equation~\ref{eqn:xray}) 
are 1 $\sigma$ errors and we use them to capture the extent to which the interacting NS-WD background is uncertain.

Given the  comoving density of galaxies, as a function of the K-band luminosity, and the 
number-luminosity distribution of LMXBs in each of these galaxies, we calculate the  specific flux received in gravitational
waves today, at a gravitational-wave frequency of $f_{\rm GW}$ as
\begin{eqnarray}
&&F_{f_{\rm GW}} = \nonumber \\
&& \int dz \frac{dV}{dz} \int dL_K \Phi(L_K) \int dL_x \frac{dN}{dL_x} \frac{L_{\rm GW}^f(L_x,z)(1+z)}{4\pi d_L^2(z)} \, ,\nonumber \\
\label{eqn:gwflux}
\end{eqnarray}
where $L_{\rm GW}^f(L_x,z)$ is the specific luminosity in gravitational waves for a binary with an accretion luminosity in X-rays, of
$L_x$, at a redshift $z$. Since interacting binaries are likely to be synchronized by tides, we will make the assumption of
circular orbits and set the eccentricity to zero. This is consistent with observationally determined small eccentricities for
millisecond pulsars (see, Benacquista 2002), which have been suggested as the end point of LMXB evolution.
 The factor of $(1+z)$ in equation~\ref{eqn:gwflux} captures the redshifting of the
frequency, via $f_{\rm GW}(z) = (1+z)f_{\rm GW}(z=0)$ while $d_L(z)=(1+z)r(z)$ is the luminosity distance when $r(z)$ is
the comoving radial distance. The comoving volume element is given by $dV = 4 \pi r(z)^2 d\rad$.

As written in equation~\ref{eqn:gwflux}, the specific flux of the GW background formed by LMXBs can be calculated based on
observationally determined parameters as long as one has a model for the GW emission from an individual binary. 
Here, following the interpretation by Bildsten \& Deloye (2004),
we assume that the LMXBs are ultra compact binaries
composed of a neutron star, with a mass 1.4 m$_{\sun}$, and a second component involving a He or C/O white dwarf.
The results derived by Bildsten \& Deloye (2004) 
can be analytically described as
\begin{eqnarray}
P_{\rm orb} &=& (10.2 \pm 1.0) {\rm mins.} \left(\frac{L_X}{10^{38} {\rm ergs}\; s^{-1}}\right)^{-0.20 \pm 0.05} \nonumber \\
m_2 &=& (0.06 \pm 0.01) m_{\sun} \left(\frac{L_X}{10^{38} {\rm ergs}\; s^{-1}}\right)^{0.21 \pm 0.05} \, 
\end{eqnarray}
in the range of 10$^{37}$ ergs s$^{-1}$ to 10$^{39}$ ergs s$^{-1}$ in the X-ray luminosity. The uncertainty here captures the range
allowed by various donor types, such as He, C and O-type white dwarfs.

While the X-ray luminosity function of point sources has been explained based on interacting NS-WD systems, this assumption may have
certain weaknesses that can affect conclusions related to our calculation. The explanation in Bildsten \& Deloye (2004) applies strictly to
point sources in elliptical galaxies while such sources have also been observed in all nearby galaxies. This modifies the overall normalization
and, in the present calculation, we have allowed for such a possibility by not specifying galaxy types over which the background is
calculated but by rather considering the total K-band luminosity function. If it turns out that the explanation, in fact,
applies only to ellipticals and the X-ray point sources in other galaxies are not related to NS-WD systems, then we may have overestimated the
background, though we do not expect this to be significant given other uncertainties in the calculation. More over, it is unlikely that the
total X-ray luminosity function is due to NS-WD systems alone. For example, there is no clear explanation for point sources with
luminosities above few times 10$^{38}$ ergs s$^{-1}$, corresponding to the Eddington-limit for accreting neutron stars.
As such sources are less in number when compared to those below the Eddington-limit, these high X-ray luminosity 
sources do not dominate the number
counts in our calculation. While these and other minor issues (such as the uncertain WD type and secondary mass) may impact the present calculation, we believe it is equally or more powerful when compared to other techniques that have been so far used to estimate the
gravitational wave background related to compact binaries. While we do not consider our estimates to be exact, considering
all uncertainties, we believe our estimates to be accurate to a factor of a few, or at most, to a factor of ten.

\section{Results \& Discussion}

The gravitational wave background calculated following the approach outlined above
is shown in Fig.~1. While we normalize the X-ray luminosity function for LMXBs
with $L_{\rm min}=10^{37}$ ergs s$^{-1}$, consistent with observational results, when calculating the background,
we extend the luminosity function to a lower value of L$_{\rm ext}=10^{34}$ ergs s$^{-1}$ to account for both the low luminosity
binaries detected in in our galaxy and nearby galaxies (see, for e.g., Grimm et al. 2002; Gilfanov 2004), and
binaries in quiescence (in Fig.~2, we vary this parameter). In Fig.~1, we plot  the specific flux (in
ergs s$^{-1}$ cm$^{-2}$ Hz$^{-1}$ (top panel) and in terms of the power spectral density $S_h(f)$ where 
$S_h = \frac{4G}{\pi c^3} \frac{F_{f_{\rm GW}}}{f_{\rm GW}^2}$;
Following the standard practice, we plot, $S_h^{1/2}$, in units of Hz$^{-1/2}$.

\begin{figure}
\centerline{\psfig{file=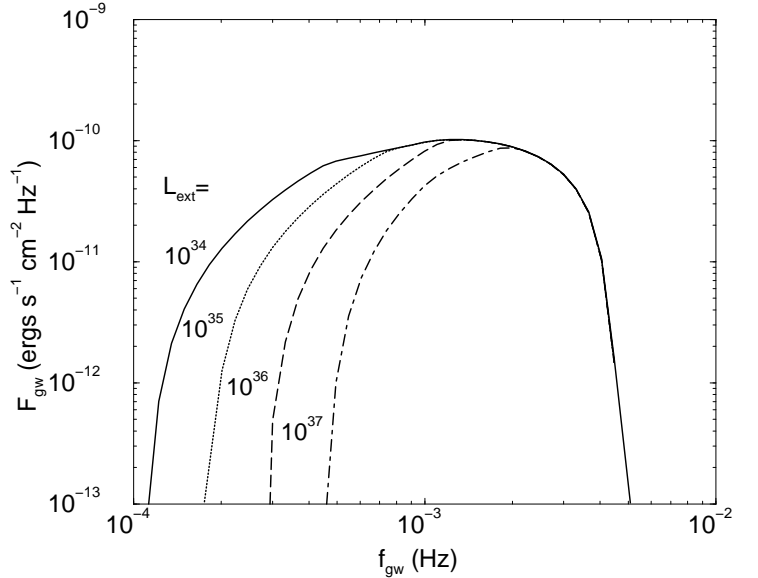,width=3.8in,angle=-90}}
\caption{The gravitational background specific flux (in units of ergs s$^{-1}$ cm$^{-1}$ Hz$^{-1}$) 
related to NS-WD binaries as a function of the X-ray luminosity (L$_{\rm ext}$ in ergs s$^{-1}$)
to which the binary X-ray luminosity function is extended. 
Here, we plot the background with L$_{\rm ext}=10^{34}$ ergs s$^{-1}$ (same as in Fig.~1 - solid line),
$10^{35}$ ergs s$^{-1}$ (dotted line), $10^{36}$ ergs s$^{-1}$ (dashed line), and
$10^{37}$ ergs s$^{-1}$ (dot-dashed line). The frequency dependence of the gravitational wave background, with varying minimum-extended luminosity, can
be understood based on the fact that orbital period increases with decreasing X-ray luminosity such that with a higher cutoff,
sources at lower frequencies disappear.}
\label{fig:xray}
\end{figure}

In calculating the background,
we make use of the fiducial parameters for the X-ray luminosity function, galaxy luminosity function and the period-mass
relations in terms of the X-ray luminosity. In Fig.~1, we have divided the contribution to redshift intervals
to show how the cosmological background  evolves both in frequency and in amplitude. While the K-band luminosity function from
Drory et al. (2003) is only out to a redshift of 1.2, we have allowed this function to scale out to a z of 3, with no changes to
the evolution; this is a safe assumption for now, though, with an increase in large scale structure observations, one can
be more certain on the fractional contribution to the gravitational wave background from the highest redshifts. In general,
at frequencies around 10$^{-3}$ the background is dominated by NS-WD binaries at redshifts less than unity. 

The spectral shape of the background, as shown in Fig.~1, is generally consistent with simple expectations: 
Following equation~1, the gravitational wave luminosity scales 
as $L_{\rm GW} \propto f^{10/3} m_c^{10/3}$, and for interacting
binaries where the mass of the donor is much less than the accretor, $m_c \propto f^{3/5}$ (Farmer \& Phinney 2003; but note the
misprint there where this is stated as $m_c \propto f^{5/3}$), and $L_{\rm GW} \propto f^{16/3}$. For the present class of
sources where the period evolution is positive, ie. the period is increasing as the mass is transferred from less massive donor
to the neutron star, the number of sources in each frequency bin roughly scales as $N(f) \propto \dot{f}^{-1} \propto m_c^{-5/3} f^{-11/3} \propto f^{-14/3}$. Since $F_{f_{\rm GW}} \propto L_{\rm GW} N(f)$, we find, $F_{f_{\rm GW}} \propto f^{2/3}$. For $z < 1$, say,
in the frequency range  between 0.5 mHz and 2 mHz, we find the slope to be roughly 0.58, in rough agreement with the expected
slope. At frequencies, below 0.5 mHz, the orbital periods are greater than an hour, and the accretion X-ray luminosity is less than 
$10^{34}$ ergs s$^{-1}$ as binaries are no longer interacting; We underestimate the background at this end of the frequency
range as X-ray observations do not capture the detached population. At frequencies above 5 mHz, the background drops
due to a sharp reduction in binaries with accretion luminosities above the Eddington limit. Here, with an increase in
the mass of the secondary donor, the number of binaries decreases. It is unlikely that we are underestimating the background
at this top end of the frequency range since interacting binaries, mainly ones already seen as LMXBs, dominate out here.

In Fig.~2, in order to understand how the gravitational wave background depends on the X-ray luminosity to which
the binaries are extended, $L_{\rm ext}$, beyond the observed minimum of the X-ray luminosity function of $L_{\rm min}=10^{37}$
ergs s$^{-1}$, we plot the specific flux as a function of $L_{\rm ext}$ between 10$^{34}$ and 10$^{37}$ ergs s$^{-1}$. Note that in all
of these cases, the total number of X-ray binaries, in each of the galaxies, is normalized by integrating the
X-ray luminosity function down to a minimum luminosity of $10^{37}$ ergs s$^{-1}$, consistent with observed results.
The difference between curves shown in Fig.~2 can be simply understood based on the fact that the orbital period is
inversely correlated with the X-ray luminosity, such that with a higher cut off in the X-ray luminosity function one 
does not include binaries with larger orbital periods or lower frequencies than with a cut off at the low end.
While luminosities of observed binaries extend down to 10$^{37}$ ergs s$^{-1}$, it is unclear whether the X-ray luminosity
function can be further extended. Expectations are such that at this low end of the luminosity range, binaries will become
transient at some critical luminosity though whether this is at 10$^{37}$ ergs s$^{-1}$ or lower is yet to be determined precisely (Bildsten
\& Deloye 2004).

In Fig.~3, we summarize the NS-WD background with respect to the LISA noise.
Here, we plot the gravitational wave amplitude $h$, which scales as $(f_{\rm GW}S_h)^{1/2}$, and have scaled the background
with an overall amplitude correction of $(f_{\rm GW} \Delta T_{\rm obs})^{-1/2}$ when accounting for the typical improvement in 
LISA noise for an observational duration of $\Delta T_{\rm obs}=1$ year. The LISA noise curve shown in Fig.~3 is for a signal-to-noise
detection of unity and follows from Larson et al. (2002). For comparison, we also show the
total background related to WDs (in orbit around other WDs or main sequence stars) as calculated by Farmer \& Phinney (2003).
To capture the uncertainty related to interacting NS-WD binaries in X-ray data, we vary the background calculation using
 parameter estimate errors in the binary distribution. The shaded region in Fig.~3 shows the extent to which background varies with
1 $\sigma$ variations (but keeping $L_{\rm ext}$ fixed at 10$^{34}$ ergs s$^{-1}$). 
The lower limit of this range should be considered as the absolute lower limit on the
expected gravitational wave background from NS-WD binaries. This is several orders of magnitude below the LISA noise
and the total WD-WD extragalactic background.

As noted before, however, we have underestimated the NS-WD background by not accounting for detached binaries.
We include these binaries based on the observed orbital period distribution of $\sim$ 80 NS-WD binaries,
related to both pulsars and LMXBs, compiled from the literature from Lorimer (2001) and Banacquista (2002).
We randomly draw from this sample to calculate the gravitational wave background of the Milky Way; While 
the period, and thus the secondary mass, distribution is established from known statistics, there is still a 
freedom with regards to  the total number of NS-WD binaries or normalization of the distribution. Taking a conservative approach to
purposely overestimate the background,
we assume that the fractional number of NS-WD binaries in the Milky Way is roughly between 3 (pessimistic) and 300 (optimistic) times that of the
number of LMXBs down to a X-ray luminosity of 10$^{34}$ ergs s$^{-1}$ (with 30 as the mid range). This results in a total number of
9000 binaries  (900 to 90000 in the whole range) with periods between roughly 200 days and 11 mins, though 
we allow the distribution to expand at both ends beyond this range. We follow the same normalization 
procedure for each galaxy, so that the previous calculation related to LMXBs alone can easily be extended.

Note that our normalization, though has a large range, is consistent with numerical estimate in the literature based on
observations (e.g., Lorimer 2001) as well population synthesis models (e.g., Nelemans et al. 2001).
Our mid total number of $\sim$ 9000 binaries suggests a birth rate of $10^{-6}$ yr$^{-1}$
in agreement with estimated lifetimes, for LMXBs, of roughly $2.5 \times 10^6$ years. The implied merger rate,
by rescaling numbers in population synthesis models of Nelemans et al. (2001) is again about $10^{-6}$ yr$^{-1}$ or 
1 Myr$^{-1}$. The 9000 number is also roughly consistent with the estimated number of NS-WD binaries, out
to a distance of 3 kpc from the Sun, of 850 (Edward \& Bailes 2001) 
if one assumes a galactic distribution with a density that scales as
${\rm sech}(z/z_0)^2 \exp (-R/R_0)$ with $z_0=200$ pc and $R_0=2.5$ kpc. Note that the overall normalization is
rather uncertain by at least to an order of magnitude.

In Fig.~3, we show the total estimated NS-WD background. Here, again, as in the case of LMXBs alone, we show
a range of estimated backgrounds by varying the total normalization by a factor of 10 both higher and lower than the mean
normalization. The estimated range suggests, however, that the total background is below that of the extragalactic
WD-WD background by at least an order of magnitude. The background is also below the LISA noise level and
is unlikely to be a major source of signal. Between 0.1 mHz and 10 mHz, we find that a spectral
slope of $\sim$ -0.8, consistent with the expectation of $(fS_h)^{1/2} \propto f^{-2/3}$ 
when the background is dominated by inspiralling detached binaries (Phinney 2001). In terms of the overall amplitude,
our estimate for the total NS-WD background is roughly compatible
with an estimate in the literature by Schneider et al. (2001), though, our calculation does not
reproduce the exact shape of their result. Their calculation suggests a bump at the low frequency end, $\sim 10^{-5}$ Hz, of the
spectrum (dominated by inspiral binaries), though we do not find such an apparent increase.
A bump would require a substantial peak in the number of NS-WD binaries with periods around few hours,
which is not observed (Willems \& Kolb 2002).

\begin{figure}
\centerline{\psfig{file=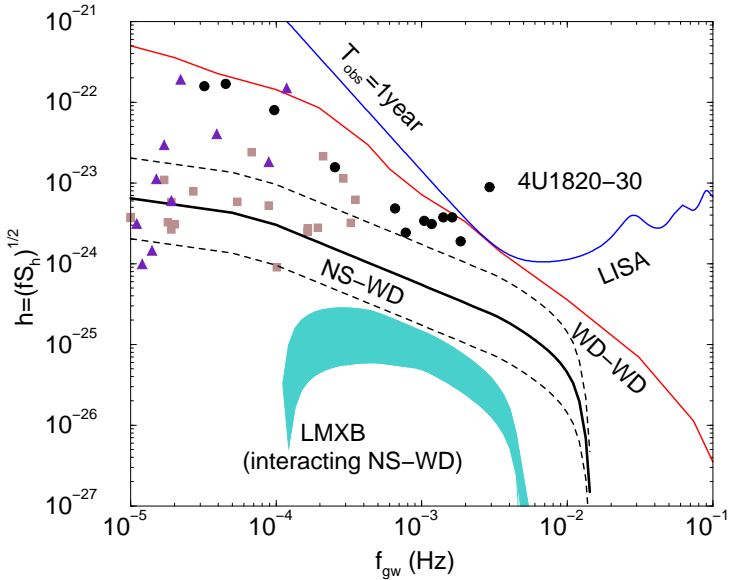,width=3.8in,angle=-90}}
\caption{The gravitational wave background related to compact binaries. Here, we compare the background generated by NS-WD
binaries with LISA noise and the total background from WD-WD binaries from Farmer \& Phinney (2003). The shaded region related to
interacting NS-WD background, based on LMXB data, captures the uncertainty associated with various parameters in the
this calculation; One can  safely consider the low end of this region as the absolute lower limit on the NS-WD background.
Note that we have underestimated the total NS-WD background  by only accounting for interacting binaries
in the LMXB phase. We account for detached binaries based on observed binary properties of NS-WD binaries
in the galaxy and based on an order of magnitude estimate for the normalization. This is rather uncertain, by at
least a factor of 10, and we show the range implied with two dotted lines; the top-most dotted curve 
shows the highest allowed total extragalactic background for all NS-WD binaries. For comparison, we also plot the
gravitational wave amplitude of several well known NS-WD binaries in our galaxy (in the case of binary millisecond pulsars, triangles show the
field binaries while squares show those that are in globular clusters; circles show LMXBs in globular clusters with
well measured orbital properties). Among these, 4U1820-30 is expected to be detected at a high signal-to-noise ratio by LISA. We estimate few 
more such sources may be detected in LISA data streams.
}
\label{fig:lisa}
\end{figure}

Our estimate on the total NS-WD background is consistent with the order of magnitude
estimate by Kim et al. (2004). There, the authors used statistics related to 3 binary pulsars
that are expected to merge within the Hubble time. The estimate in Kim et al (2004) is roughly
an order of magnitude below the LISA noise and also below the extragalactic background related to
WD-WD binaries. In comparison,  the normalization related to NS-WD binaries is roughly consistent between these two studies
(9000 here vs. a total of 10400 there), though the number density of galaxies
in our calculation is lower (with $\phi_*=0.01$ compared to 0.17). 
This difference, however, is absorbed by other uncertain parameters in our calculation.

While the extragalactic background may not be detected with LISA, the galactic population may be
within its reach. In Fig.~3, we show the gravitational wave amplitude for the sample of known 
NS-WD binaries  as used in describing the total background. 
Under the assumption that the primary mass is
a neutron star with a mass of 1.4 m$_{\sun}$ we calculate  the amplitude of the gravitational wave as
\begin{equation}
h=1.5 \times 10^{-21} \left(\frac{f_{\rm GW}}{mHz}\right)^{2/3}\left(\frac{1\; {\rm kpc}}{d}\right)\left(\frac{m_c}{m_{\sun}}\right)^{5/3} \, .
\end{equation}
We also show the amplitude for a subsample of LMXBs with well measured periods. 
 Currently, 
there are $\sim$ 100 LMXBs known in the galaxy, while 13 are present in globular clusters 
(Benacquista 2002).
Among the total sample of $\sim$ 80 binaries, including binary millisecond pulsars, we find 4U1820-30
to be the only source that may be  detectable with LISA. We can scale our results based on
published estimates in the literature. With a merger rate of $1.4 \times 10^{-4}$ yr$^{-1}$ and a total
number of NS-WD binaries 2.2 $\times 10^6$, Nelemans et al. (2001) found 38 resolved systems in LISA (at
frequencies above $\sim$ mHz), while a total of 124 were found to be detected over the whole frequency
range when they also included the total galactic background at the low frequency end. Lowering these
numbers to a merger rate of 10$^{-6}$ yr$^{-1}$, we find 0.2 resolved systems and 1 detectable system.
If we allow for the merger rate of  $10^{-5}$ yr$^{-1}$ as advocated by Kim et al. (2004), $\sim$ 10
systems  may be detected with LISA; it is unlikely that the number expected for LISA is substantially higher than this.

 Assuming typical signal-to-noise levels of order 5, we find that the
locations of these binaries will be determined by LISA to an accuracy of few degrees (Cooray et al. 2004).
Since the binary population is overly represented in globular clusters, the counterparts may
easily be determined by searching for periodic signatures, consistent with gravitational
wave data, in globular clusters within the error boxes. The detectable NS-WD sample could
potentially aid in understand the evolutionary properties of these binaries within our galaxy
which, in return, could help to establish the exact background related to these binaries at
cosmological distances. Another application of combined gravitational wave and electromagnetic radiation observations of
such binaries include extracting of fundamental physical parameters such as the graviton mass as discussed in
Cutler et al. (2002) with regards to the potential LISA observations of 4U1820-30 (also, Cooray \& Seto 2004).

\section{Summary}

Here, we discussed the stochastic background of gravitational waves from ultra compact neutron star-white dwarf (NS-WD) binaries
at cosmological distances. The background is calculated by avoiding population synthesis approaches in the literature.
We, however, make the
assumption that accreting neutron stars and donor white dwarf stars  form
most of the low mass X-ray binaries (LMXBs); this assumption is consistent with a variety of
observations that associate pulsars with LMXBs as well arguments based on binary evolution properties.
Our calculation makes use of recent results related to 
the luminosity function determined from X-ray observations for such binaries.  

The background is substantially lower;
The calculation based on LMXBs, however, does not include the total population of NS-WD binaries as we neglect detached binaries
that are not interacting as well as a large sample of binary millisecond pulsars. We extend our calculation to include
such binaries based on the observed distribution of binary periods and secondary masses, though we have a large uncertainty with
respect to an overall normalization. Even after allowing for a conservatively high number of detached NS-WD binaries,
at frequencies between 10$^{-5}$ Hz and 10$^{-1}$ Hz, as relevant for the
Laser Interferometer Space Antenna (LISA), we find that the NS-WD background is at least 
an oder of magnitude below that due to extragalactic white dwarf-white dwarf binaries. The signal is, again, 
below the detectability level of LISA. While the extragalactic background is unlikely to be detectable, we suggest 
that a few galactic NS-WD binaries will be resolved with LISA in addition to the known source
4U1820-30. 

{\it Acknowledgments:} 
We thank Alison Farmer for WD-WD background information and Naoki Seto for useful discussions.
We also thank the referee, Vicky Kalogera, for useful comments and for correspondances regarding the Kim et al. (2004) paper.
This work is supported by the Sherman Fairchild foundation and DOE DE-FG 03-92-ER40701.

\end{document}